\documentclass[12pt]{iopart}
\usepackage{graphicx}
\usepackage{iopams}
\usepackage{epsfig}
\usepackage{bm}

\begin{document}
\title[One-dimensional molecular spin chains]{Probing magnetic order
  and disorder in the one-dimensional
  molecular spin chains  CuF$_{2}$(pyz) and $[Ln$(hfac)$_{3}$(boaDTDA)$]_{n}$
  ($Ln=$Sm, La) using implanted muons}
\author{T. Lancaster$^{1}$, B.M. Huddart$^{1}$, R.C. Williams$^{1,2}$, F. Xiao$^{1,3,4}$,
  K.J.A. Franke$^{1,5,6}$, P.J. Baker$^{7}$, F.L. Pratt$^{7}$,
  S.J. Blundell$^{8}$, J.A. Schlueter$^{9,10}$, M.B. Mills$^{11}$ A.C. Maahs$^{11}$ and K.E. Preuss$^{11}$}
\address{$^{1}$Durham University, Centre for Materials Physics, Durham, DH1 3LE, United Kingdom}
\ead{tom.lancaster@durham.ac.uk}
\address{$^{2}$Department of Physics, University of Warwick, Gibbet
  Hill Road, Coventry CV4 7AL, United Kingdom}
\address{$^{3}$Department of Chemistry and Biochemistry, University of
  Bern, Freiestrasse 3, CH-3012 Bern, Switzerland}
\address{$^{4}$Laboratory for Neutron Scattering and Imaging, Paul Scherrer Institut, CH-5232 Villigen PSI, Switzerland}
\address{$^{5}$School of Physics and Astronomy, University of Leeds,
  Leeds LS2 9JT, United Kingdom}
\address{$^{6}$Department of Materials Science and Engineering, University of California, Berkeley, Berkeley, California 94720, USA}
\address{$^{7}$ISIS Facility, Rutherford Appleton Laboratory, Chilton, Didcot, 
  OX11 0QX, United Kingdom}
\address{$^{8}$Department of Physics, University of Oxford, Clarendon
  Laboratory, Parks Road, Oxford, OX1 3PU, United Kingdom}
\address{$^{9}$Materials Science Division, Argonne National Laboratory,
  Argonne, Illinois 60439, USA}
\address{$^{10}$Division of Materials Research, National Science
  Foundation, 2415 Eisenhower Ave, Alexandria, VA 22314, USA}
\address{$^{11}$Department of Chemistry, University of Guelph, Guelph, Ontario N1G 2W1, Canada}

\begin{abstract}
We present the results of muon-spin relaxation ($\mu^{+}$SR)
measurements on antiferromagnetic and ferromagnetic spin chains.
In antiferromagnetic CuF$_{2}$(pyz) we identify a transition to long
range magnetic order taking place at $T_{\mathrm{N}} = 0.6(1)$~K,
allowing us to estimate a ratio with the intrachain exchange of $T_{\mathrm{N}}/|J| \approx 0.1$ and the ratio of
interchain to intrachain exchange 
coupling as $|J'/J| \approx 0.05$. The ferromagnetic chain
[Sm(hfac)$_{3}$(boaDTDA)]$_{n}$ undergoes an ordering transition at
$T_{\mathrm{c}}=2.8(1)$~K, seen via a broad freezing of dynamic fluctuations on the muon
(microsecond) timescale and implying $T_{\mathrm{c}}/|J| \approx 0.6$. The ordered radical moment continues to fluctuate on
this timescale down to 0.3~K, while the Sm moments remain
disordered. In contrast, the radical spins in [La(hfac)$_{3}$(boaDTDA)]$_{n}$ remain magnetically
disordered down to $T=0.1$~K suggesting $T_{\mathrm{c}}/|J| < 0.17$. 
\end{abstract}
	\submitto{\JPCM}

\section{Introduction}
A major theme in the investigation of molecular magnets has concerned materials whose
low-energy physics arises from interactions that are constrained
to act in less than three spatial dimensions \cite{molmagreview_1, molmagreview_2}. This is achieved by 
linking localized spins using molecular ligands in such a way
as to promote strong interactions between the 
magnetic centres only along particular directions, with weaker ones
acting along others. Particularly notable is the one-dimensional (1D) spin
chain, since an idealised version of this problem can be thoroughly described theoretically,
revealing a rich range of correlations and an exotic spectrum of excitations \cite{giamarchi}.
Although an isolated one dimensional spin chain will not show any
long-range magnetic order \cite{sachdev}, 
real materials will host interactions between chains, however weak,
which have a strong effect on their properties. 

A quasi-one-dimensional Heisenberg magnet with principal exchange
interaction $J$ acting along the chains \cite{note} and a smaller
interaction $J'$ linking the chains is described by the Hamiltonian \cite{mu_review3}
\begin{equation}
\hat{H} = J \sum_{\langle ij\rangle}
\hat{\boldsymbol{S}}_{i}\cdot\hat{\boldsymbol{S}}_{j}
+J' \sum_{\langle ij'\rangle}
\hat{\boldsymbol{S}}_{i}\cdot\hat{\boldsymbol{S}}_{j'}, 
\label{model2}
\end{equation}
where the sum $\langle ij \rangle$ is over nearest-neighbour spins on
the same chain and $i$ and $j'$ label spins on adjacent chains. 
In order to probe 1D behavior in a system described by this model, we
work in a temperature regime 
where $|J'| \ll  T \ll |J|$. When we approach a regime
where the characteristic energy of the thermal fluctuations allows
detail at the scale of
$J'$ to be resolved, we should expect that the three-dimensional nature of the
material  becomes evident and it will 
magnetically order. However, the strongly anisotropic nature of the
exchange ($|J'/J|\ll 1$) will have consequences on the ordered
state that arises \cite{sengupta1} and the ordering temperature is driven
downwards by the 1D nature of the system compared to a
three-dimensional one.  
The ordered magnetic moments will also be
renormalised \cite{schulz}, making them potentially very small, and difficult to
observe with magnetic susceptibility measurements. Finite, but possibly
very long, correlation lengths may exist above the ordering temperature, 
reducing the entropy in the system. Consequently, the entropy change
on ordering will be much reduced compared to a more isotropic system and
may prevent specific heat measurements from detecting a
transition \cite{sengupta1}. For this reason sensitive magnetic measurement techniques
such as muon-spin relaxation ($\mu^{+}$SR) have found a use in detecting the transitions
to long range magnetic order that allow us to quantify the strength of
the interchain exchange and reveal how well real materials are
described by idealised models \cite{anotherdim, mu_review3,ben}

\begin{figure}
\begin{center}
  \includegraphics[width=6cm]{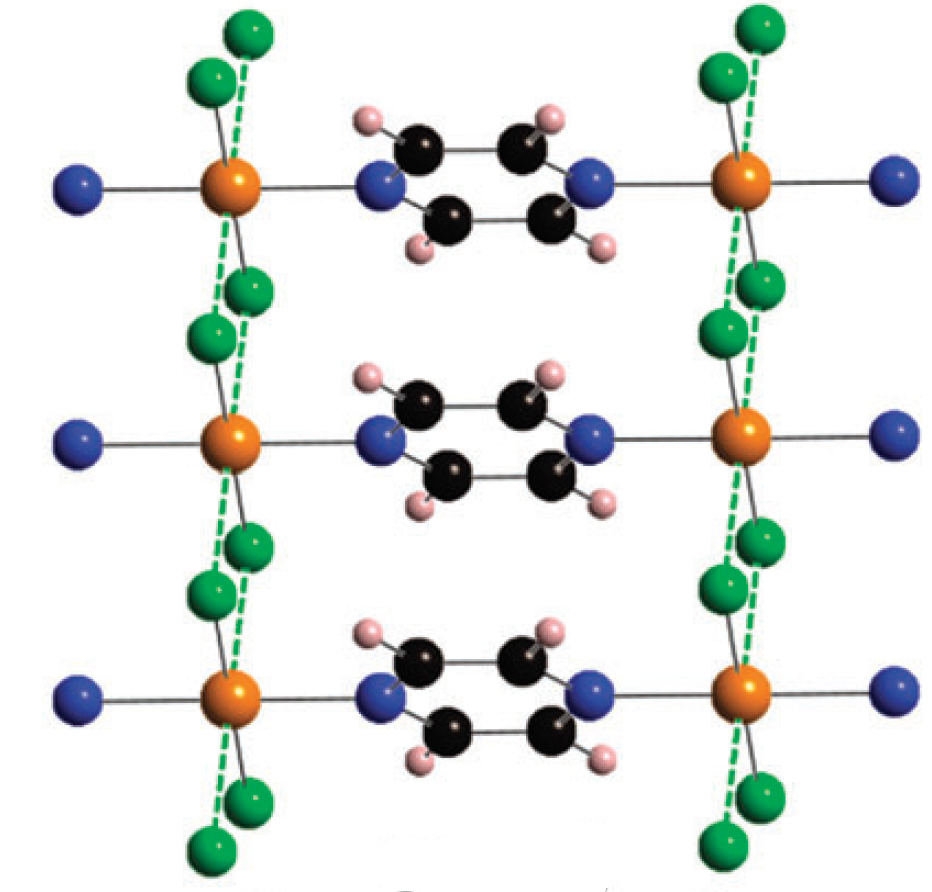}\\
  \includegraphics[width=\columnwidth]{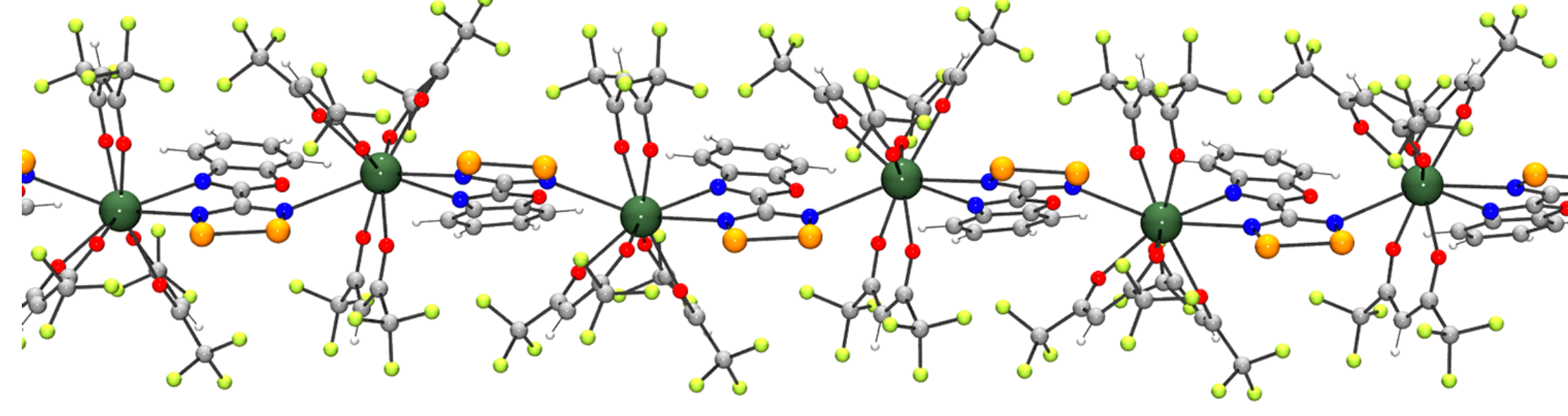}
\caption{{\it Top}: structure of CuF$_{2}$(pyz) with the strong
  antiferromagnetic coupling along the horizontal Cu$-$pyz$-$Cu
  chains \cite{lapidus} (black, C; blue N; pink H, orange Cu; green F). {\it Bottom}:
  structure of [$Ln$(hfac)$_{3}$(boaDTDA)]$_{n}$, where ferromagnetic coupling
  is mediated along the chains of boaDTDA radical ligands via the
  $Ln$(hfac)$_{3}$ units \cite{sm}. (dark green, $Ln$; orange, S; red,
  O; grey, C; white, H; blue, N; pale green,
  F.)\label{fig:structures}} 
\end{center}
\end{figure}

In this paper we present $\mu^{+}$SR measurements on two different 1D
coordination polymer 
systems: (i) the quasi-1D Heisenberg antiferromagnet CuF$_{2}$(pyz), built
from Cu$^{2+}$ spins linked antiferromagnetically with pyrazine (pyz =
C$_{4}$H$_{4}$N$_{2}$) ligands [Fig.~\ref{fig:structures} (top)] \cite{lapidus};
(ii) a quasi-1D Heisenberg ferromagnetic series based on a
$-[Ln^{3+}$-radical$]_{n}-$ paradigm, namely
[$Ln$(hfac)$_{3}$(boaDTDA)]$_{n}$  [Fig.~\ref{fig:structures} (bottom)]
where $Ln=$Sm, La;  (boaDTDA) is the radical ligand
4-(benzoxazol-20-yl)-1,2,3,5-dithiadiazolyl and  hfac =
1,1,1,5,5,5-hexafluoroacetylacetonato \cite{la,sm}. In both systems,
the muon relaxation is dominated by the magnetism of the nuclear spins
in the high-temperature
paramagnetic regime owing to the rapid fluctuations of the electronic
spins on the timescale of muon spin precession.
However, on cooling, as fluctuations in the electronic degrees of
freedom slow and enter the muon time window, the nature of the muon
coupling with the system changes and transitions to regimes of
magnetic order are detected. 

\section{Experimental}

In a $\mu^{+}$SR experiment \cite{steve} spin-polarized
positive muons are stopped in a target sample, where the muon usually
occupies an interstitial position in the crystal.
The observed property in the experiment is the time evolution of the
muon spin polarization, the behaviour of which depends on the
local magnetic field at the muon site. Each muon decays, with an average 
lifetime of 2.2~$\mu$s, into two neutrinos and a
positron, the latter particle being
emitted preferentially along
the instantaneous direction of the muon spin.
Recording the time dependence of the positron emission directions
therefore allows the determination of
the spin-polarization of the ensemble of muons.
In our experiments, positrons are
detected by detectors placed forward (F) and
backward (B) of the initial muon polarization direction.
Histograms $N_{\mathrm{F}}(t)$ and $N_{\mathrm{B}}(t)$ record the number of
positrons detected in the two detectors as a
function of time following
the muon implantation. The quantity of interest is
the decay
positron asymmetry function, defined as
\begin{equation}
A(t)=\frac{N_{\mathrm{F}}(t)-\alpha_{\mathrm{exp}} N_{\mathrm{B}}(t)}
{N_{\mathrm{F}}(t)+\alpha_{\mathrm{exp}} N_{\mathrm{B}}(t)} \, ,
\end{equation}
where $\alpha_{\mathrm{exp}}$ is an
experimental calibration constant. $A(t)$ is proportional to the
spin polarization of the muon ensemble.

Zero-field (ZF) muon-spin relaxation ($\mu^+$SR) measurements 
were carried out on polycrystalline samples using the MuSR and HIFI spectrometers
at the STFC-ISIS Facility, Rutherford Appleton Laboratory UK. For the
measurements on CuF$_{2}$(pyz),
a sample was packed in an Ag foil envelope (foil thickness
12.5$\mu$m), and 
attached to the cold finger of a $^{3}$He cryostat.
For measurements on [$Ln$(hfac)$_{3}$(boaDTDA)]$_{n}$ samples were
sealed in an Ag foil envelope inside an airtight sample holder and
mounted on the cold finger of a dilution refrigerator.

\section{One-dimensional antiferromagnetism in Cu$X_{2}$(pyz)}
The isostructural series Cu$X_{2}$(pyz), where $X = $Br, Cl, F, share a rectangular
lattice structure composed of $S = 1/2$ Heisenberg Cu$^{2+}$ ions. The Cu$^{2+}$ ions are
linked into chains by bridging pyrazine molecules, which are themselves cross-linked
via dihalide bridges to create 2D layers (Fig.~\ref{fig:structures}, top). Antiferromagnetic
superexchange pathways are therefore provided by both pyrazine 
and dihalide bridges, and interlayer magnetic coupling is small
in comparison. For both $X=$Br and Cl materials, the largest exchange
strength $J=46$~K (Br) and $=28$~K (Cl) was suggested to act along the halide bridges, with
smaller exchange of around 8~K along the pyz chains \cite{butcher}.  Previous $\mu^{+}$SR investigations have determined transition
temperatures for $X = $Br and Cl \cite{tom}, yielding ratios of the
ordering temperature to principal coupling of
$T_{\mathrm{N}}/|J| \approx 0.08$ (Br) and $\approx 0.11$ (Cl),  
although the significant value of the Cu-pyz-Cu coupling means that
these systems are best regarded as rectangular lattice antiferromagnets.
This coordination polymer family has more recently been extended to include the
$X =$ F system, CuF$_{2}$(pyz) \cite{lapidus} which is notable for the
small value of exchange coupling promoted along the fluorine bridges,
estimated to be of order 1~K, while the principal exchange is along
the Cu$-$pyz$-$Cu links and was estimated to be $J = 7$~K, such that
quasi-1D behaviour might be expected, with no magnetic transition
identified down to $T\approx 0.6$~K. 

\begin{figure}
\begin{center}
  \epsfig{file=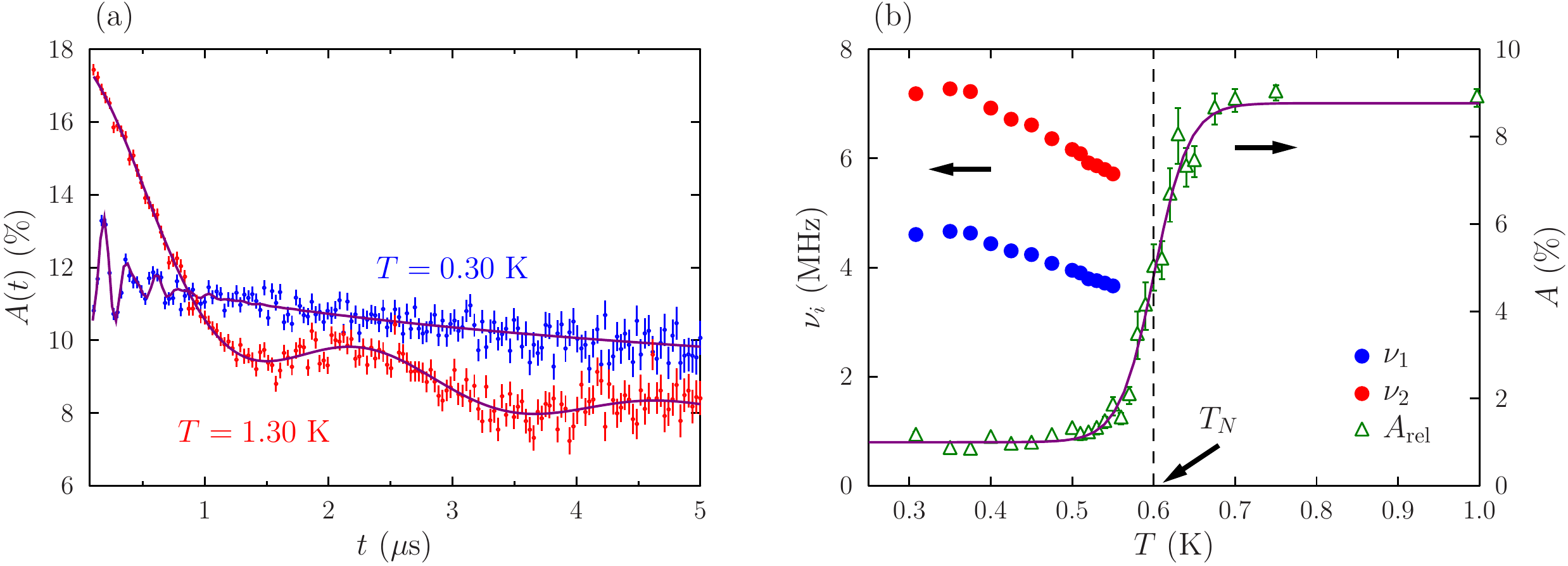, width=\columnwidth}
\caption{(a) ZF data measured on CuF$_{2}$(pyz) above and below
  $T_{\mathrm{N}}$.
  (b) Temperature evolution of ({\it left axis}) precession frequencies
  and ({\it right axis}) relaxing amplitude $A_{\mathrm{rel}}$.\label{fig:schl1}}
\end{center}
\end{figure}

Example ZF  $\mu^{+}$SR data for CuF$_{2}$(pyz) are shown in Fig.~\ref{fig:schl1}(a),
where oscillations are seen at all measured temperatures. 
 The low frequency oscillations for $T>0.6$~K are common in
 fluorine-containing materials and are caused by
dipole-dipole coupling between the muon and fluorine nuclei
\cite{fmuf,fmuf3}.  These oscillations are
typically resolved in the paramagnetic regime of molecular magnets of
this type since, in this regime, the electronic moments on the Cu$^{2+}$ ions fluctuate
rapidly on the muon (microsecond) timescale and are motionally
narrowed from the spectrum, leaving the quasistatic nuclear moments to
relax the muon spins \cite{fmuf2}.
The data were found to be best
described by the oscillations characteristic of the bound state of a $\mu^{+}$ and a single
F ion, and were fitted to a function of the form
\begin{equation}
A(t) = A_{1}D(t)e^{-\lambda_{1}t} + A_{2}\exp^{-\sigma^{2}t^{2}}
 + A_{\mathrm{bg}}e^{-\lambda_{\mathrm{bg}}t}, 
\label{eq:fits_rob}
\end{equation}
where the first term represents the contribution from muons strongly
coupled to fluorine nuclei, the second to muons coupled to a
distribution of other quasistatic nuclei  and the third term accounts
for muons that implant in the sample holder. 
In Eqn~(\ref{eq:fits_rob}), $D(t)$ accounts for the muon-fluorine
coupling, which takes the form \cite{fmuf}
\begin{equation}
  D(t) = \frac{1}{6}
  \left[
1 + 2\cos\left(\frac{\omega t}{2}\right) + \cos(\omega t) +
2\cos\left(\frac{3\omega t}{2}\right)
  \right],
\end{equation}
where $\omega = \mu_{0}\gamma_{\mu}\gamma_{\mathrm{F}}\hbar/4\pi r^{3}$,
$\gamma_{i}$ are gyromagnetic ratios and
$r$ is the muon-fluorine distance. 
The fitted value of
$\omega =1.74(2)~\mu$s$^{-1}$ corresponds to a muon-fluorine separation of $r = 1.09(1)$~\AA.
(This value represents a time-averaged bond length, as the frequencies of vibrational
modes exceed those accessible by $\mu^{+}$SR.) It is therefore likely
that one class of muon sites occurs near the fluorine nuclei and
contributes the component with amplitude $A_{1}$ in Eqn~(\ref{eq:fits_rob}), while
another couples closer to the pyz groups, giving rise to the component
with amplitude $A_{2}$. This is consistent with the behaviour usually
seen in materials of this type \cite{steele} and with computations of
the muon sites using electronic structure methods \cite{xiao}. 

The more rapid oscillations measured for $T<0.5$~K are
characteristic of a quasistatic local magnetic field at the 
muon stopping site caused by long range magnetic order \cite{steele}.
This causes a coherent precession of the
spins of those muons with a component of their spin polarization
perpendicular to this local field. The frequencies of the oscillations are given by
$\nu_{i} = \gamma_{\mu} B_{i}/2 \pi$, where $\gamma_{\mu}$ is the muon
gyromagnetic ratio ($=2 \pi \times 135.5$~MHz T$^{-1}$) and $B_{i}$
is the average magnitude of the local magnetic field at the $i$th muon
site. This is indicative of long-range magnetic
order of the Cu$^{2+}$ ions, and also implies the existence of two magnetically inequivalent
muon stopping sites.
Asymmetry spectra were fitted to damped cosinusoidal functions
[$A_{i}{\rm e}^{-\lambda_{i}t}\cos(2\pi\nu_{i}t+\phi_{i})$, where $\phi_{i}$ is a phase
offset and $\lambda_{i}$ a relaxation rate] and the temperature dependence of the two resolvable
precession frequencies are shown in Fig.~\ref{fig:schl1}(b), where their relative magnitudes were held in
fixed proportion.
As was found in the $X=$Br and Cl materials
\cite{tom}, the oscillations become heavily damped and therefore unresolvable as the magnetic
transition is approached from below. 
This effect means that frequencies cannot be reliably fitted close to
the N\'{e}el temperature $T_{\mathrm{N}}$. Instead,
data were binned heavily (in order to smooth out the oscillations) and fitted to a
purely decaying asymmetry relaxation function
$A_{\mathrm{rel}}\exp(-\lambda t)$. A significant change is visible in the
relaxing amplitude $A_{\mathrm{rel}}$ [Fig.~\ref{fig:schl1}(b)], which
allows us to estimate $T_{\mathrm{N}} = 0.6(1)$~K.
(This step change in relaxing amplitude reflects a fraction of lost asymmetry
within the ordered state, where large, static or slowly 
fluctuating internal magnetic
fields lead to rapid precession outside of the resolution limit set by
the muon pulse width, and therefore tracks the evolution of static
internal fields in the sample.)

A useful figure of merit in comparing the success with which a material
realizes one-dimensionality is $T_{\mathrm{N}}/|J|$, as this
quantity should be zero in the ideal case and close to unity for an
isotropic material. This quantity may also be
used to estimate $J'$, which is the parameter of interest in the
Hamiltonian in Eqn~(\ref{model2}). One method of doing this in
antiferromagnetic chains has been developed from
Quantum Monte Carlo computations \cite{yasuda}, whose results may be approximated 
using the expression
\begin{equation}
|J'|/k_{\mathrm{B}} =\frac{T_{\mathrm{N}}}{
4 c \sqrt{  
\ln\left( 
\frac{a|J|}{k_{\mathrm{B}}T_{\mathrm{N}}}
\right)
+
\frac{1}{2}\ln \ln
\left(
\frac{a |J|}{k_{\mathrm{B}}T_{\mathrm{N}}}
\right)
}},
\label{eq:yasuda}
\end{equation}
with $a=2.6$ and $c=0.233$. 
This yields a value of $|J'/J| \approx 0.05$ (assuming $J=7$~K), suggesting it is a
reasonable approximation of a 1D Heisenberg antiferromagnet,
comparable in the isolation of its chains with KCuF$_{3}$ 
\cite{mu_review3}\footnote{Note that there is an error in the expression in
  Eqn~(\ref{eq:yasuda}) in Refs.~\cite{molmagreview_1} and
  \cite{mu_review3}, where the square root in the denominator has been
omitted.}. The exchange $J'$ represents an effective geometric
average over the three-dimensional exchange couplings in the lattice,
but which will be dominated in this case by the coupling along the fluorine
bibridges. As a result, this value is consistent with the estimate of
1~K coupling though these links made from the susceptibility data \cite{lapidus}.
 In addition, coupled chain mean field theory \cite{schulz} predicts an ordered
moment of $\approx$0.2~$\mu_{\mathrm{B}}$, consistent with the
muon precession frequencies measured. 

\section{One-dimensional ferromagnetism in
  [$Ln$(hfac)$_{3}$(boaDTDA)]$_{n}$, $Ln=$Sm, La}

The $-$[(metal)$^{3+}-$(radical)]$-$ series
[$Ln$(hfac)$_{3}$(boaDTDA)]$_{n}$, was the first reported coordination polymer including a
bridging DTDA (-1,2,3,5-dithiadiazolyl) radical ligand \cite{la}. 
The structure is formed from $Ln^{3+}$(hfac)$_{3}$ units arranged along
the crystallographic $b$-direction, linked with $S=1/2$ radical
boaDTDA ligands (Fig.~\ref{fig:structures}, bottom). 
The 1D chains pack such that the heteroatoms of
the boaDTDA ligands do not allow any close contacts
between chains \cite{la,sm}.
Materials with $Ln=$La and Sm were both found to show 1D ferromagnetic
exchange interactions between the radical spins, mediated via the
$Ln^{3+}$ ions \cite{la,sm}. 
Such FM exchange can be achieved if there is a
different symmetry between the magnetic orbital of each
radical and the orbitals of the bridging species with which it overlaps. This is the case
here owing to the
$\approx 30^{\circ}$ angle between the planes of the radical species
on each side of the bridging ion \cite{la}.

Results of room temperature magnetometry  in the $Ln=$La material \cite{la}  are consistent with a
diamagnetic La$^{3+}$ metal ion ($4f^{0}$) and one $S=1/2$ spin per radical unit. 
Low temperature susceptibility measurements 
showed FM interactions which have principal ferromagnetic exchange
strength of $J=-0.58$~K.
The dc magnetic susceptibility of the $Ln=$Sm material \cite{sm} is consistent 
with the expected behaviour of one
Sm$^{3+}$ metal ion ($4f^{5}$, $g_{J}J=0.71$) and one 
$S_{\mathrm{rad}} = 1/2$ boaDTDA radical ($g_{J} J = 1$) per repeat
unit. 
Again, low temperature dc susceptibility suggests
ferromagnetic interactions with exchange constant  $J =-4.4(2)$~K. 
Notably, below $T=3.1$~K, the magnetic susceptibility in the $Ln=$Sm material becomes
strongly magnetic field-dependent, suggesting the stabilisation of a
ferromagnetically ordered state.
The magnetic order was suggested to involve the larger radical
spins only, with the Sm$^{3+}$ spins apparently remaining disordered down to $T=2$~K.

The FM interchain coupling in the Sm system was explained
using the McConnell I mechanism \cite{sm,mcconnell}.
Here FM coupling between net magnetic
moments on large molecules can occur via 
 the contact between $\alpha$ (positive, up) spin density
on one molecule and $\beta$ (negative, down) spin density on its
neighbour.
When the dominant contact is between $\alpha$ and $\beta$ it results in antiparallel alignment
of these moments, and therefore alignment between the majority
$\alpha$ components and overall FM coupling.

\begin{figure}
\begin{center}
\epsfig{file=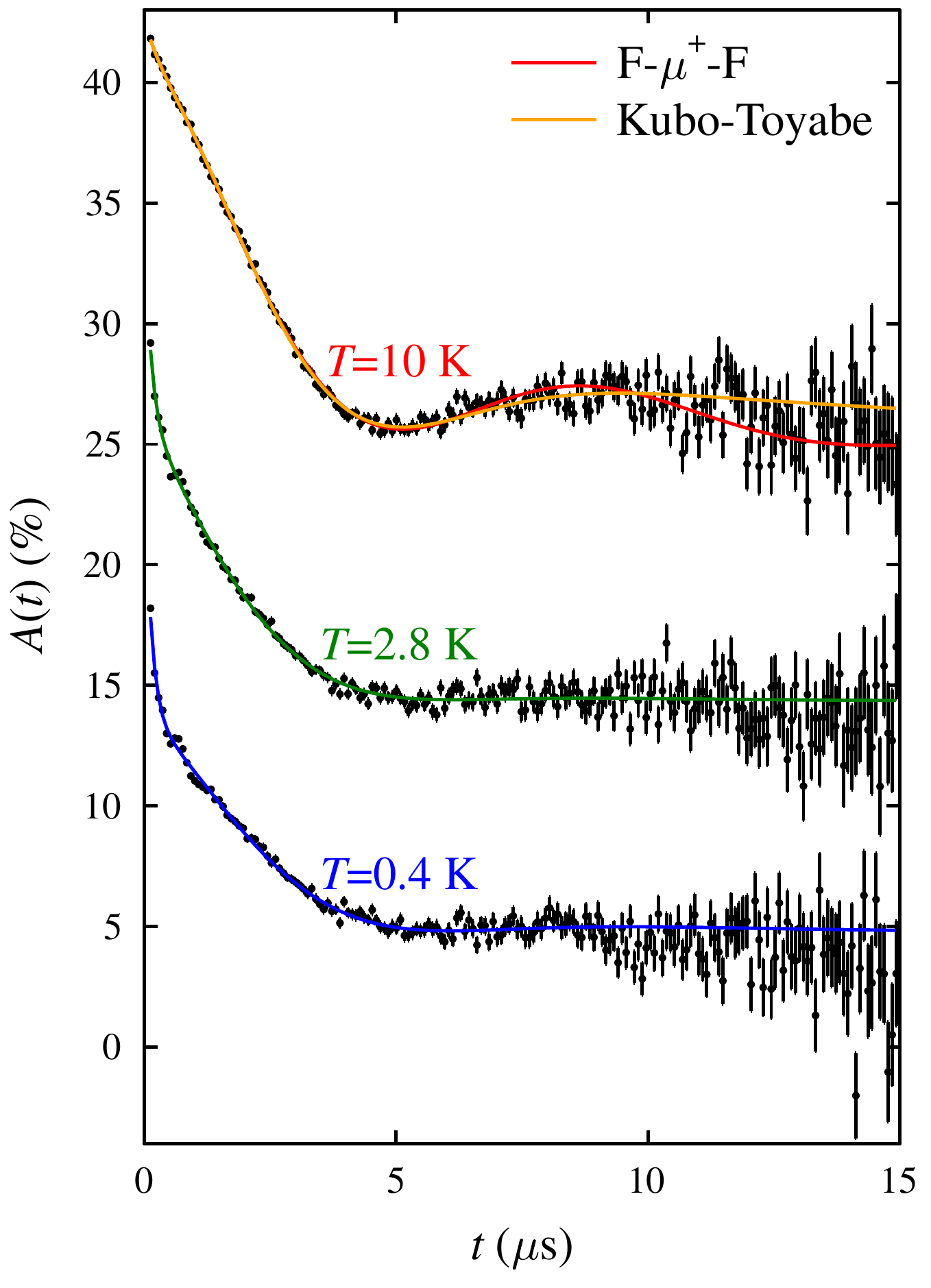, width=7cm}
\caption{Example spectra measured on the $Ln$=Sm material at three
  different temperatures. [Spectra offset for clarity by $A(t)+10\%$ for $T=10$~K
  and $T=2.8$~K.] Lines are fits explained in the text.\label{fig:sm_dat}}
\end{center}
\end{figure}

ZF $\mu^{+}$SR measurements of the $Ln$=Sm material were made at
temperatures
between $T=0.2$~K and 20~K.  Example spectra measured at three different
temperatures are shown in Fig.~\ref{fig:sm_dat}.
 The
 spectra at  temperatures $>4$~K exhibit a low frequency oscillation that
 looks intermediate in character between the F-$\mu^{+}$
 oscillations seen for CuF$_{2}$(pyz) and a Kubo-Toyabe (KT) function.
 The KT function [$D(t) = \frac{1}{3} + \frac{2}{3}(1-\Delta^{2}
 t^{2})\exp(-\Delta^{2}t^{2}/2)$]
 results from the coupling of muons to a dense, classical distribution
of random, static moments with magnitudes centred around zero, which gives rise to a variance in the
magnetic field distribution at the muon
site of $\Delta^{2} = \gamma_{\mathrm{\mu}}^{2}\langle B^{2} \rangle$. In contrast, the muon-dipole coupling discussed
above models the quantum mechanical interaction of a small number of
moments. The intermediate nature of the relaxation seen here, is
consistent with the muon spin being relaxed by several nuclear moments localized
close to the muon. (Numerical simulations of dipole-dipole interactions are
consistent with this scenario and contrasts with the case of
relaxation from a dilute array of moments, which gives rise to a
Lorentzian relaxation \cite{uemura}.) Given the presence of
(electronegative) fluorine in the hfac
groups, muon sites in their vicinity might be expected, while the
chemical complexity of the material makes it plausible that the muon spin
will also be relaxed by coupling to other nuclei in its vicinity. 

In addition to the relaxation from nuclear spins,  a rapidly relaxing component (with relaxation
 rate $\lambda_{2}$) is observed at early times
 which becomes steadily more prominent as the temperature is
 lowered. 
As a result,  we fit the data to a function of
the form     
\begin{equation}
A(t)=A_{\mathrm{rel}} [p_{1} e^{-\lambda_{1} t} D(t)+p_{2}
e^{-\lambda_{2} t} ]
+A_{\mathrm{bg}},\label{eq:sm_fit}
\end{equation}
where $D(t)$ models the muon coupling to the local nuclei,
$\lambda_{i}$ are relaxation rates and $p_{i}$ give the proportion of
the signal arising from each component. In addition to the relaxing
asymmetry $A_{\mathrm{rel}}$ there is also a contribution
$A_{\mathrm{bg}}$ from those muons that do not relax, including those that implant in the sizeable sample
holder. 
We first fitted the high temperature spectra to a  F-$\mu^{+}$-F model
\cite{fmuf} (Fig.~\ref{fig:sm_dat}, red line) and obtained a
characteristic frequency $\omega/2\pi=0.046$~MHz, corresponding to
large $\mu^{+}$-F distances of $r\approx 2.0$~\AA.  These spectra are
also well captured by a KT model (Fig.~\ref{fig:sm_dat}, orange line)
with $\Delta=0.34$~$\mu$s$^{-1}$, which is fairly typical for
relaxation due to nuclear moments in the paramagnetic regime of a
molecular material.  As the $\mu^{+}$-F distance required by the
F-$\mu^{+}$-F model is unusually large compared to the typical bond
length of $\approx 1.1$~\AA, the low temperature data were fitted to
Eqn~(\ref{eq:sm_fit}) with $D(t)$ described by the KT model, with the field width fixed to its average value $\Delta=0.29$~$\mu$s$^{-1}$.  The relaxing amplitude and rapid relaxation rate in Eqn~(\ref{eq:sm_fit}) were also found not to vary significantly with temperature and so were fixed to their average values $A_{\mathrm{rel}}=20.1$\%, and rapid relaxation rate $\lambda_{2}=7.4$~$\mu$s$^{-1}$.

\begin{figure}
\begin{center}
\epsfig{file=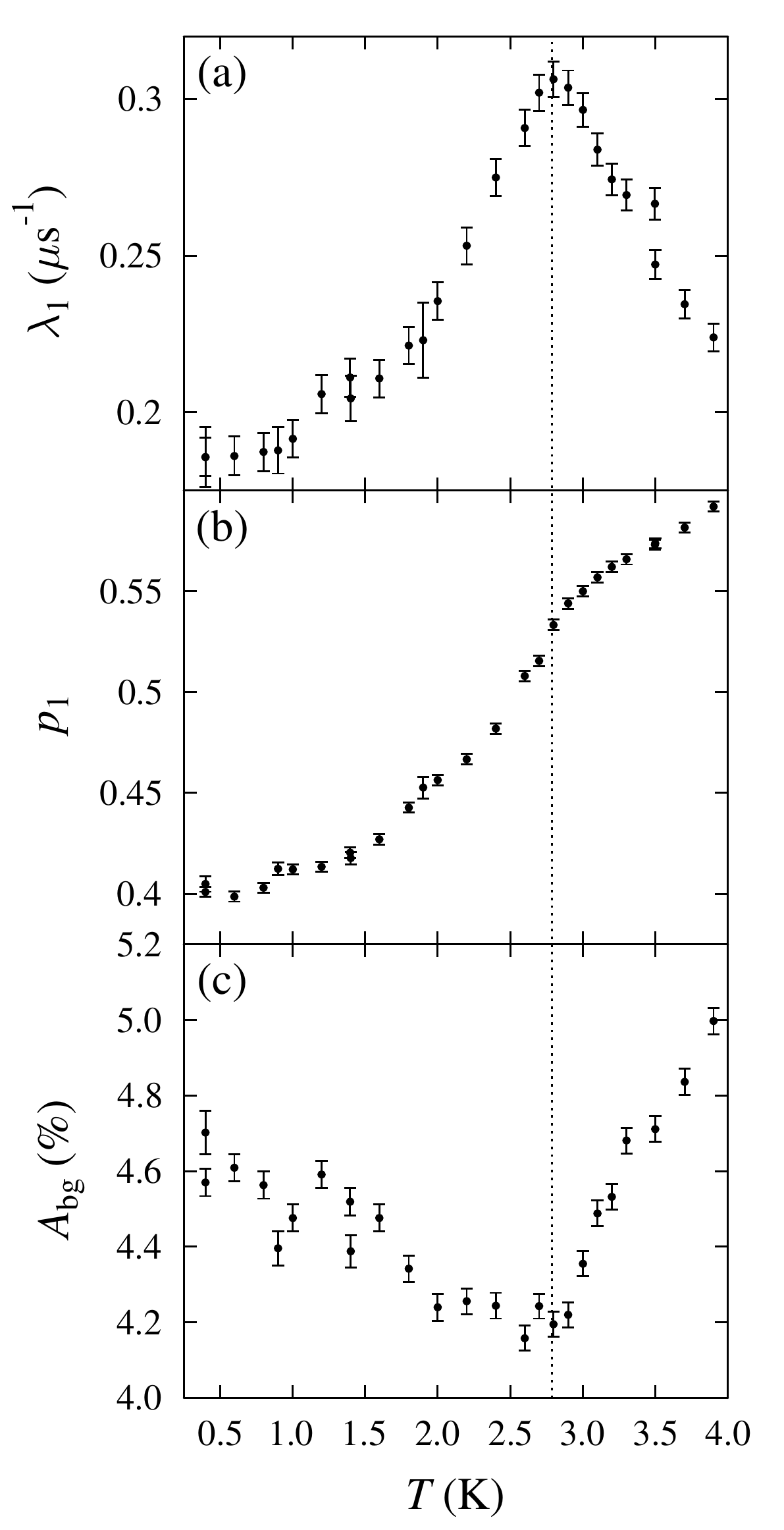, width=7cm}
\caption{Fitted parameters from fits to Eqn~(\ref{eq:sm_fit}) for the
  $Ln=$Sm material. (a) Slow
  relaxation rate $\lambda_{1}$, (b) slow relaxing proportion $p_{1}$
  and (c) baseline contribution $A_{\mathrm{bg}}$. \label{fig:sm_fit}}
\end{center}
\end{figure}

The behaviour of the fitted parameters is shown in
Fig.~\ref{fig:sm_fit}. 
A sharp peak in the smaller relaxation rate 
$\lambda_{1}$  is seen at $T_{\mathrm{c}}=2.8$~K [Fig.~\ref{fig:sm_fit}(a)].  This
roughly coincides with the magnetic ordering
transition reported at $T=3$~K previously
from magnetic susceptibility measurements \cite{sm}, and so we ascribe the
residual relaxation of the KT function to the fluctuations of the radical
spins. 
Previous muon measurements made on molecular radical-based magnets
have seen oscillations in the muon asymmetry accompanying transitions
to long range
magnetic order \cite{blundell2} at frequencies that can be resolved
using the pulsed muon beam as the ISIS facility. It is therefore
likely that the fact that oscillations are not evident in this case
reflects a more complex magnetic state below the transition
temperature. 
In the fast fluctuation limit, we expect the relaxation rate $\lambda_{1}$ to vary as
$\lambda_{1} = 2 \Delta^{2} \tau$, where $\Delta$ is the width of the
distribution of magnetic fields ($\Delta^{2} = \gamma_{\mu}^{2}\langle B^{2}  \rangle$) and $\tau$ is its correlation time. The peak is therefore
consistent with a slowing of the radical moments as the transition is
approached from above. The lack of discontinuity at the transition
suggests that the moments continue to fluctuate below
$T_{\mathrm{c}}$. 
We also see a minimum in $A_{\mathrm{bg}}$ in this temperature region
[Fig.~\ref{fig:sm_fit}(c)]. In addition to muons that stop outside the
sample, this component also has a contribution from those muons that
lie along the direction of local static fields (expected to be 1/3 of
the total in a polycrystalline sample). The minimum in $A_{\mathrm{bg}}$ is
seen at $T_{\mathrm{c}} = 2.8$~K and is consistent with slow fluctuations
relaxing the largest proportion of muons at this temperature. As the
temperature is reduced further, the slow recovery of $A_{\mathrm{bg}}$
suggests that fewer muons are relaxed by the slow dynamics,
consistent with a spatially inhomogeneous freezing out of the
dynamics. 
This is also
consistent with the decrease in the proportion of muons $p_{1}$ relaxed by the
radical spins, which decreases smoothly as temperature is lowered. 

Our data therefore show clear evidence for a bulk magnetic
transition at $T=2.8$~K, which suggests a ratio $T_{\mathrm{c}}/|J| = 0.63$.
However, the lack of oscillations and the behaviour of the dynamic
relaxation
suggests that  the $Ln=$Sm ferromagnetic transition seen
previously involves significant dynamic fluctuations in the muon
spin-precession time
window, which is typically sensitive to correlation times of order
$\tau\approx 0.1-10$~ns. This is also consistent with the reported ac
susceptibility measurements, where a response was noted in the
out of phase component $\chi''$ close to 3~K \cite{sm}. We might
therefore speculate that the ferromagnetically ordered state likely
has a finite correlation length below $T_{\mathrm{c}}$, with
fluctuations over a broad range of timescales freezing out in a
spatially inhomogeneous manner as the
temperature is lowered. 

Also notable in this material is the fast relaxation component with
relaxation rate $\lambda_{2}$. This varies very little as temperature
is lowered. As such relaxation is not seen in the $Ln=$La material
(see below) we ascribe it to relaxation from the Sm$^{3+}$ moments,
which appear to fluctuate down to the lowest measured temperature of
0.3~K. (Such low-temperature fluctuations
likely involve low-energy
transitions between the $^{6}H_{5/2}$ levels of the Sm$^{3+}$ ion's
ground state, split by the crystal fields into three Kramers doublets
with the potential for further splitting resulting from the local magnetic field on the ordered radical spins.)
Since we constrain $p_{1}+p_{2}=1$, we see the number of muons
relaxed by these spins increases as temperature is lowered and the
radical spin fluctuations are frozen out.

\begin{figure}
\begin{center}
\epsfig{file=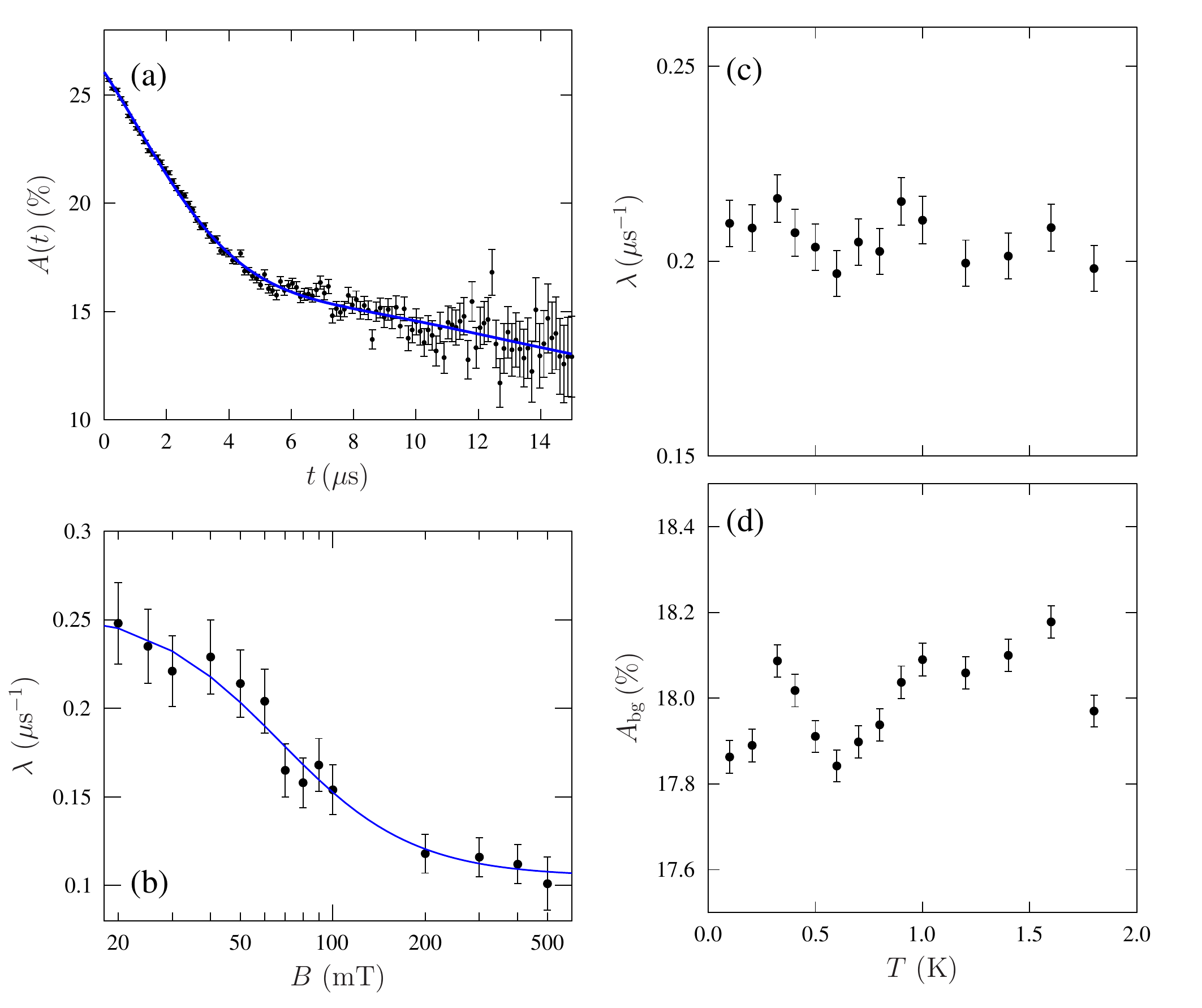, width=0.85\columnwidth}
\caption{(a) Example spectra measured on the $Ln=$La material at
  $T=0.1$~K. (b) Longitudinal field measurements made at $T=1.5$~K. 
  The blue line shows a fit to a Redfield function. 
(c) Relaxation rate  and (d) baseline $A_{\mathrm{bg}}$ for fits of
Eqn~(\ref{eq:la_material}) to the data. 
  \label{fig:la_data}}
\end{center}
\end{figure}

For measurements made on the $Ln=$La material, 
little difference is observed in the spectra in the measured
temperature range $0.1 \leq T \leq 20 $~K. 
Example data are shown in Fig.~\ref{fig:la_data}(a).  In this case no
fast relaxing component is seen, consistent with our assignment of it
being due to magnetism of the metal ion in the $Ln=$Sm material.
There is also less structure in the slow relaxation, which is more
consistent with a damped KT function, suggesting that 
those muons that are relaxed by the Sm$^{3+}$ ions in the $Ln=$Sm
material are relaxed by nuclei (and residual radical spin
fluctuations) in the $Ln=$La material, leading to a broader
distribution of magnetic fields at the muon site. 
We fitted the data with the
function
\begin{equation}
A(t) = A_{\mathrm{rel}} e^{-\lambda t} D(t) + A_{\mathrm{bg}},
\label{eq:la_material}
\end{equation}
where $D(t)$ is the Kubo-Toyabe function with
$\Delta=0.24$~$\mu$s$^{-1}$. 

As we do not see an appreciable change in relaxation over the measured
temperature range [Fig.~\ref{fig:la_data}(c)], it is unlikely that
there is any magnetic transition in the measured temperature range. This would suggest
that $T_{\mathrm{c}}/|J|<0.17$, making this a potentially successful realization of
a 1D ferromagnetic chain material \cite{f4bimnn}.
There is a subtle change in the apparent background $A_{\mathrm{bg}}$, as shown in 
Fig.~\ref{fig:la_data}(d), where we see that $A_{\mathrm{bg}}$ decreases
below 1~K reaching a minimum at around 0.6~K, and a
local maximum at 0.3~K. The size of these variations is small, but 
as they take place around the energy scale of the
exchange constant $J$ and so might reflect freezing of relaxation
channels of individual spins. However, our main result is that,
unlike for the $Ln=$Sm material, there is not evidence for collective
ordering of the magnetic radical spins. 

In order to gain further insight into the magnetism in the
disordered $Ln$=La material,
measurements were made as a function of magnetic field,
applied longitudinal to the initial muon spin direction. 
The relaxation rates as a function of applied field $B$ at fixed
temperature $T=1.5$~K are
shown in Fig.~\ref{fig:la_data}(b). At this temperature ($T\gg |J|$) we would
expect the excitations to be single spin flips. The data are well
described by a Redfield model for which $\lambda =
\Delta^{2}\tau/[1 + (\gamma_{\mu} B\tau)^{2}]$ \cite{hayano}, appropriate for a dense array of
dynamically fluctuating spins, with a characteristic fluctuation time
$\tau \approx 35$~ns and a field width of $\Delta/\gamma_{\mu} \approx
7$~mT (the latter being fairly typical of radical magnets probed by
implanted muons \cite{blundell2,blundell3}). This is consistent with fluctuations of the radical spins,
which appear to persist to the lowest measured temperatures and it is
likely these parameters also describe the electronic fluctuation distribution of
radical spins in the $Ln=$Sm material.

\section{Conclusions}
We have presented the results of muon spin relaxation measurements on
two different molecular spin chain systems. In CuF$_{2}$(pyz) we
observed a transition to a regime of long-range magnetic order below
$T=0.6(1)$~K and have shown that the system represents a reasonably
successful realization of a quasi 1D Heisenberg antiferromagnet. The
radical magnet system [$Ln$(hfac)$_{3}$(boaDTDA)]$_{n}$ with $Ln=$Sm and
La represents one of the most chemically complex magnetic systems
investigated using $\mu^{+}$SR to date. We have  shown that the
magnetic transition in the $Ln=$Sm material likely represents a
freezing of slow dynamics giving rise to a dynamically fluctuating and
spatially inhomogeneous magnetic state
for $T<T_{\mathrm{c}}$ with Sm$^{3+}$ spins continuing to fluctuate
down to the lowest temperatures. It is unclear whether the driver for
these slow magnetic
fluctuations is the array of Sm$^{3+}$ spins or is related to the
extended nature of the electron density distribution over the
radical units that results in the McConnell coupling mechanism.
We note that this is not generic behaviour for a 1D ferromagnetic
chain of radical spins, as several examples have been previously
studied and show conventional magnetic order at low temperature \cite{blundell2,sugano,f4bimnn}.
The $Ln=$La material shows no sign of
magnetic order down to $T=0.1$~K, with the radical spins remaining in a dynamically fluctuating
magnetic state. 

\ack
This work was carried out at the STFC-ISIS facility, Rutherford
Appleton Laboratory, UK and we are
grateful for the provision of beamtime. We thank EPSRC, UK and
STFC for financial support. The Natural Science and Engineering Research Council (NSERC) of Canada supported KEP's contribution to this work through a Discovery Grant (DG).
JAS acknowledges support from the Independent Research/Development
(IRD) program while serving at the National Science Foundation. 
Research data from this publication will be made
available via Durham University. ({\it DOI(s) will be inserted here.})


\begin{thebibliography}{10}
\bibitem{note}
{We use the
single-$J$ convention here where the energy gap between the singlet and
triplet level of a two spin system with Hamiltonian $\hat{H} = J
\hat{\boldsymbol{S}}_{i}\cdot\hat{\boldsymbol{S}}_{j}$
is given by $J$. }
\end{thebibliography}
\end{document}